\documentclass[10pt,twocolumn]{article}
\usepackage[margin=0.75in]{geometry}
\usepackage[T1]{fontenc}
\usepackage{lmodern}
\usepackage{amsmath,amssymb}
\usepackage{booktabs,tabularx,array}
\usepackage{tikz}
\usetikzlibrary{arrows.meta,positioning}
\usepackage{enumitem}
\usepackage{xurl}
\usepackage[numbers,sort&compress]{natbib}
\usepackage[hidelinks]{hyperref}
\usepackage[stretch=30,shrink=20]{microtype}
\usepackage{placeins}

\setlist{nosep,leftmargin=*}
\newcolumntype{Y}{>{\raggedright\arraybackslash}X}

\title{DiffSynth-Music: Audio-Conditioned KV-Cache Adapters\\
for Controllable Music Generation}
\author{
    Zhongjie Duan\textsuperscript{1}, Shengchuan Gao\textsuperscript{2},
    Hong Zhang\textsuperscript{1}, Yingda Chen\textsuperscript{1}\\[0.5em]
    {\small \textsuperscript{1}Alibaba Group\quad
    \textsuperscript{2}Shanghai Jiao Tong University}\\[0.25em]
    {\small \href{mailto:duanzhongjie.dzj@alibaba-inc.com}{\nolinkurl{duanzhongjie.dzj@alibaba-inc.com}}}\\
    {\small \href{mailto:gscdy111@sjtu.edu.cn}{\nolinkurl{gscdy111@sjtu.edu.cn}}}
}
\date{}

\begin{document}
\maketitle

\begin{abstract}
Text and lyrics specify broad musical characteristics and sung content but offer limited control over musical timing, melody, and reference-based style. We introduce DiffSynth-Music\footnote{Models: \url{https://modelscope.cn/models/DiffSynth-Studio/DiffSynth-Music}}, a framework that adds composable audio conditioning to a music synthesis backbone through layer-wise key-value injection. The three template models, Control, Prosody, and Reference, are initialized from the backbone diffusion transformer and trained with conditional flow matching. They support five control types: beats, vocals, accompaniment, prosody, and reference audio. A shared variational autoencoder maps conditioning waveforms into a common latent space, enabling their attention memories to be combined. With the template timestep fixed at the clean-data endpoint and other inputs held constant, each control cache is computed once and reused throughout sampling. Training pairs are derived from music recordings using beat extraction, source separation, vocal resynthesis, and reference-excerpt selection. Single-control evaluations on Mandarin and English songs demonstrate improved adherence across all five control types and better lyric fidelity under vocal conditioning relative to the backbone. Automatic music-quality and instruction-following scores remain broadly comparable to those of the evaluated base models, with metric-specific trade-offs. We release the three template models to support research and creative applications in controllable music generation.
\end{abstract}

\section{Introduction}
\label{sec:introduction}

Recent advances in music generation have improved the ability of generative models to synthesize music from natural-language descriptions and lyrics. MusicGen~\cite{musicgen}, Stable-Audio-Open~\cite{stableaudio}, and the ACE-Step family~\cite{acestep,acestep15} explore different approaches to translating semantic descriptions into musical audio. Full-song generation systems, including MiniMax-Music3~\cite{minimax}, DiffRhythm-2~\cite{diffrhythm}, HeartMuLa~\cite{heartmula}, SongBloom~\cite{songbloom}, and LeVo-2~\cite{levo2}, further address lyric alignment, musical coherence, and generation quality. Together, these developments provide a foundation for generating songs whose broad musical characteristics and sung content are specified through text.

Specifying a song through text, however, differs from controlling its musical realization. A prompt can describe genre, instrumentation, or mood, while lyrics specify the words to be sung; neither directly determines the timing of individual beats, the progression of a vocal melody, or the phrasing of a performance. For tasks that require following an existing musical idea, these details are more directly conveyed through audio. This motivates extending text- and lyrics-conditioned generators with audio-based control, allowing users to guide not only the overall character of a song but also specific aspects of its performance.

Prior work~\cite{musiccontrolnet,jasco} introduces time-varying conditions to complement global text descriptions. Audio references provide such temporal information while also conveying acoustic characteristics that are difficult to specify verbally. Temporally aligned signals can constrain the progression of a composition, whereas a short excerpt can guide style, vocal delivery, and timbre without prescribing an output timeline. Integrating these complementary forms of guidance within a pretrained generator is the central objective of this work.

We propose \textbf{DiffSynth-Music}, an audio-conditioned adapter initialized from a pretrained music-generation DiT. The adapter transforms control-audio latents into attention keys and values that are injected into the generation branch. This design introduces reference-dependent attention memory while preserving the backbone's text-conditioning and waveform-decoding pathways.

The framework supports five complementary control types. Beats encode rhythmic events; vocals provide melodic and performance information; accompaniment specifies instrumental context; prosody represents vocal pitch and timing with reduced pronunciation cues; and Reference supplies global stylistic context from a short excerpt. Their shared representation also permits joint conditioning. In particular, accompaniment and prosody can constrain instrumental structure and vocal melody without requiring preservation of the original vocal timbre. Such flexibility is a design objective whose realization must be assessed separately from basic target adherence.

The contributions of this work are as follows:
\begin{itemize}
    \item \textbf{Architecture.} We propose DiffSynth-Music, a DiT-based audio-conditioning architecture that injects layer-wise key-value representations into a pretrained generation backbone. A shared waveform-to-latent interface unifies five control types and enables their composition through joint attention memory.
    \item \textbf{Open-source models.} Using conditional flow matching, we train and open-source three audio-conditioned template models: Control, Prosody, and Reference. Control supports beat, vocal, and accompaniment control, while Prosody and Reference provide vocal pitch and timing guidance and reference-based stylistic conditioning, respectively. All three models can be combined at inference.
    \item \textbf{Experimental validation.} Experiments on Mandarin and English song generation demonstrate that DiffSynth-Music improves adherence across all five audio-control types while maintaining automatic music-quality and instruction-following scores broadly comparable to those of the evaluated base models.
\end{itemize}

\section{Related Work}
\label{sec:related}

\subsection{Diffusion and Flow Matching}

Diffusion and flow-based models provide a foundation for continuous generative modeling. Denoising diffusion probabilistic models~\cite{ddpm} learn to reverse a gradual noising process, while latent diffusion~\cite{latentdiffusion} moves generation into a pretrained autoencoder's representation space. Diffusion transformers (DiTs)~\cite{dit} replace convolutional denoisers with transformers operating on latent tokens. These developments motivate the use of latent representations and transformer backbones for music synthesis.

Flow matching~\cite{flowmatching} learns continuous-time vector fields by regressing velocities along prescribed probability paths, without simulating the generative dynamics during training. Its formulation accommodates both diffusion-derived and alternative transport paths. Rectified flow~\cite{rectifiedflow} learns transport along straight-line interpolations between source and target samples. DiffSynth-Music uses a linear noise-to-data path and augments the resulting conditional velocity model with audio-derived attention memory, rather than introducing a new generative objective.

\subsection{Music Generation Architectures}

Music-generation systems differ in how they represent audio and coordinate long-range structure with local acoustic detail. MusicGen~\cite{musicgen} autoregressively models discrete audio tokens under text and melody conditioning. YuE~\cite{yue} extends token-based modeling to long-form lyrics-to-song generation, while HeartMuLa~\cite{heartmula} combines an audio-token language model with conditioning on musical descriptions and lyrics. These approaches use sequence modeling to organize musical content over time. Stable-Audio-Open~\cite{stableaudio} and ACE-Step~\cite{acestep} use latent diffusion architectures; ACE-Step-1.5~\cite{acestep15} combines language-model planning with a DiT. DiffRhythm-2~\cite{diffrhythm} uses semi-autoregressive block flow matching to address lyric--vocal alignment. MiniMax-Music3~\cite{minimax} combines hierarchical language modeling with flow-matching synthesis from fused language-model hidden states. SongBloom~\cite{songbloom} interleaves autoregressive musical sketching and diffusion refinement. LeVo-2~\cite{levo2} first models mixed musical tokens for semantic planning, then predicts vocal and accompaniment tokens in parallel before diffusion-based waveform reconstruction.

These systems explore complementary choices of representation, generation order, and acoustic synthesis. DiffSynth-Music instead focuses on extending an existing generator: it uses ACE-Step-1.5-XL-SFT~\cite{acestep15} as its backbone and adds composable audio-conditioned templates while keeping the generation backbone frozen.

\subsection{Controllable Music Generation}

Controllable generation complements global descriptions with conditions that specify musical content more directly. MusicGen~\cite{musicgen} supports melody conditioning, and Music-ControlNet~\cite{musiccontrolnet} introduces time-varying melody, dynamics, and rhythm controls into spectrogram diffusion. JASCO~\cite{jasco} uses flow matching to combine text with symbolic and audio conditions, including chords, melodies, and drum references. Its information bottlenecks and temporal blurring selectively preserve information relevant to each condition. Reference-based song generation is also supported by YuE~\cite{yue}, HeartMuLa~\cite{heartmula}, and SongBloom~\cite{songbloom}, illustrating that audio prompting is not restricted to diffusion architectures.

DiffSynth-Music builds on these directions by representing five audio-control types through a shared waveform-to-latent interface and combining their layer-wise attention memories. To reduce pronunciation cues, DiffSynth-Music constructs its prosody condition through pitch- and envelope-based vocal resynthesis rather than through the bottleneck and temporal-blurring strategy used by JASCO. The focus is on composing temporally aligned musical constraints and global reference guidance within a frozen pretrained DiT, not on introducing audio conditioning itself.

\subsection{Adapting Pretrained Generators}

Pretrained generators can be adapted through additional conditioning branches, attention context, or constrained parameter updates. ControlNet~\cite{controlnet} connects a trainable branch to a frozen diffusion backbone through zero-initialized convolutions. T2I-Adapter~\cite{t2iadapter} learns lightweight modules that translate control inputs into features for a frozen text-to-image model. LoRA~\cite{lora} instead represents trainable weight updates with low-rank factors. These methods provide distinct mechanisms for adapting generation without retraining all backbone parameters.

Attention-based adaptation is particularly relevant to our design. Prefix-tuning~\cite{prefix} supplies learned task-specific attention context, whereas IP-Adapter~\cite{ipadapter} derives conditioning from input images and uses separate text and image cross-attention pathways. DiffSynth-Music likewise derives context from each conditioning input, but uses DiT-initialized audio templates to produce layer-wise keys and values. These are concatenated with the generation branch's keys and values within joint attention, rather than processed through IP-Adapter's decoupled cross-attention. This distinction separates our conditioning interface from both fixed learned prefixes and residual feature injection.

\subsection{Music Analysis}

Music analysis provides tools for constructing audio-derived control signals. Essentia~\cite{essentia} includes multi-feature and reliability-informed beat trackers~\cite{multifeaturebeat,degarabeat}. Hybrid-Transformer-Demucs~\cite{demucs} combines waveform- and spectrogram-domain representations for source separation, while pYIN~\cite{pyin} estimates fundamental-frequency trajectories. In our framework, these tools support beat, source, and prosody extraction. Their outputs are estimated annotations and signals, so extraction errors can affect the resulting conditioning data.

Beyond task-specific signal analysis, multimodal large language models such as Qwen3-Omni~\cite{qwen3omni} support audio understanding and can generate textual annotations from audio inputs. These capabilities complement beat tracking, source separation, and pitch estimation by providing semantic descriptions and transcriptions of vocal content. In our data-annotation pipeline, Qwen3-Omni supplies descriptive prompts and lyric annotations for the evaluation recordings, linking the musical audio to the textual conditions used for generation.

\section{Methodology}
\label{sec:method}

\subsection{Problem Formulation}

Let $\mathbf{x}$ denote a target music waveform and $y$ its textual condition, including the prompt and lyrics. The set of control identifiers is $\mathcal{M}=\{\mathrm{beats},\mathrm{vocals},\mathrm{accompaniment},\mathrm{prosody},\mathrm{reference}\}$. A nonempty subset $\mathcal{S}\subseteq\mathcal{M}$ identifies the active conditions, and $\boldsymbol{c}_{\mathcal{S}}=\{\mathbf{c}_m\}_{m\in\mathcal{S}}$ denotes their waveforms. Single-condition generation corresponds to $|\mathcal{S}|=1$. The objective is to synthesize a waveform that is consistent with the text and with the musical attributes specified by the active conditions.

The backbone consists of a text encoder $T_\psi$, a VAE encoder-decoder pair $(E_\omega,D_\omega)$, and a generation DiT $F_\theta$, where $\theta$ denotes the pretrained parameters of the base DiT and remains frozen during template training. This decomposition follows latent-space generative modeling~\cite{vae,latentdiffusion,dit} and its audio applications~\cite{stableaudio,acestep15}. The text encoder produces the text representation $\mathbf{h}_y$, while the shared VAE encoder maps the target and control waveforms to the clean target latent $\mathbf{z}_1$ and control latents $\mathbf{z}_{c,m}$, respectively. The encoder $E_\omega$ includes the audio preprocessing required by the backbone. Sharing the VAE places target and control waveforms in a common latent space. Throughout the paper, $t\in[0,1]$ denotes dimensionless flow time, with $t=0$ denoting the noise endpoint and $t=1$ denoting the data endpoint. The generated terminal latent is decoded as $\widehat{\mathbf{x}}=D_\omega(\widehat{\mathbf{z}}_1)$.

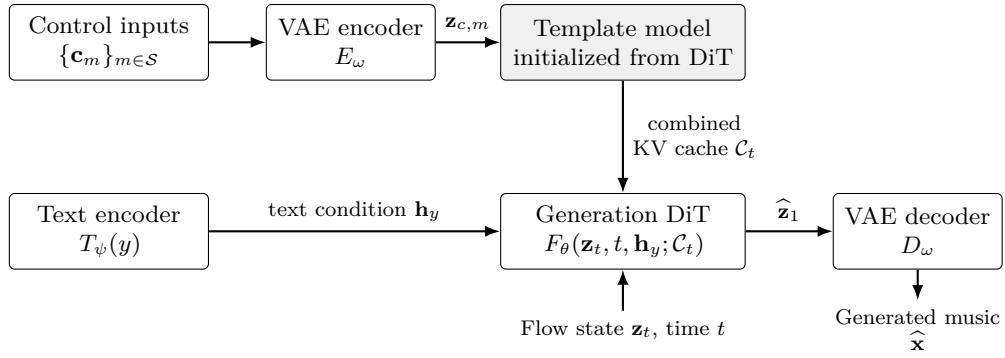
\begin{figure*}[t]
\centering
\begin{tikzpicture}[
    block/.style={draw,rounded corners=2pt,align=center,minimum height=1cm,font=\small},
    flow/.style={-{Latex[length=2mm]},thick},
    note/.style={font=\footnotesize,align=center}
]
    \node[block,text width=2.4cm] (reference) at (0,2.5) {Control inputs\\$\{\mathbf{c}_m\}_{m\in\mathcal{S}}$};
    \node[block,text width=2cm] (encoder) at (3.2,2.5) {VAE encoder\\$E_\omega$};
    \node[block,text width=3cm,fill=gray!12] (adapter) at (6.8,2.5) {Template model \\initialized from DiT};
    \node[block,text width=2.4cm] (text) at (0,0) {Text encoder\\$T_\psi(y)$};
    \node[block,text width=3cm] (generator) at (6.8,0) {Generation DiT\\$F_\theta(\mathbf{z}_t,t,\mathbf{h}_y;\mathcal{C}_t)$};
    \node[block,text width=2cm] (decoder) at (10.7,0) {VAE decoder\\$D_\omega$};
    \node[note] (output) at (10.7,-1.3) {Generated music\\$\widehat{\mathbf{x}}$};
    \node[note] (state) at (6.8,-1.3) {Flow state $\mathbf{z}_t$, time $t$};
    \draw[flow] (reference) -- (encoder);
    \draw[flow] (encoder) -- node[above,note] {$\mathbf{z}_{c,m}$} (adapter);
    \draw[flow] (adapter) -- node[right,note] {combined\\KV cache $\mathcal{C}_t$} (generator);
    \draw[flow] (text) -- node[above,note] {text condition $\mathbf{h}_y$} (generator);
    \draw[flow] (state) -- (generator);
    \draw[flow] (generator) -- node[above,note] {$\widehat{\mathbf{z}}_1$} (decoder);
    \draw[flow] (decoder) -- (output);
\end{tikzpicture}
\caption{DiffSynth-Music conditions a pretrained generation DiT through audio-derived KV representations. The shared VAE and adapter process each active control, and their keys and values are combined at matched attention blocks. The connection from the DiT to the decoder summarizes the integration of the learned vector field from noise at $t=0$ to the generated latent at $t=1$.}
\label{fig:architecture}
\end{figure*}

Figure~\ref{fig:architecture} summarizes the architecture. Each active control waveform is encoded by the shared VAE and processed by the adapter; the resulting key-value representations condition the generation branch. The following formulation combines concatenation-based attention fusion with adapter-only optimization.

\subsection{DiT-Based Template Modules}

The architecture of DiffSynth-Music is built on Diffusion-Templates~\cite{diffusiontemplates}, a unified plugin framework that separates base-model inference from the injection of controllable capabilities. Within this framework, we organize the five audio control modes into three separately parameterized template modules: \textbf{Control (C)}, \textbf{Prosody (P)}, and \textbf{Reference (R)}. Control handles beat, vocal, and accompaniment inputs; Prosody processes resynthesized vocals that preserve pitch and timing; and Reference processes short music excerpts for stylistic guidance. Thus, the five modes describe the conditioning signals, whereas C, P, and R identify the three modules that process them. Multiple modes can be activated together through the shared KV-injection interface; the construction of these signals is detailed in Section~\ref{sec:controls}.

This three-module organization is an empirical design choice rather than a theoretically prescribed partition of the five modes. We do not assume that this is the optimal partition among all possible module organizations.

Each template module $A_{\phi_j}$ follows the backbone DiT architecture, where $j\in\{C,P,R\}$ indexes the three modules and $\phi=(\phi_C,\phi_P,\phi_R)$ collects their parameters. The parameters of each module are initialized from pretrained DiT weights. The template modules and generation branch maintain separate parameters with compatible hidden-state and attention-head dimensions; the generation branch retains the frozen pretrained parameters $\theta$. For control type $m$, let $g(m)\in\{C,P,R\}$ identify the corresponding template module. The selected module processes $\mathbf{z}_{c,m}$, while the generation branch processes the current flow state $\mathbf{z}_t$.

For an attention block $\ell$ and one attention head, let $\mathbf{H}^{A,m}_{\ell}\in\mathbb{R}^{L_{\ell,m}\times d_{\ell}}$ denote the adapter states associated with control $m$, where $L_{\ell,m}$ is the token count and $d_{\ell}$ is the hidden width. Head indices and dependence on auxiliary inputs are suppressed for readability. Keys and values are obtained through the module-specific attention projections~\cite{transformer}.
\begin{equation}
    \begin{aligned}
        \mathbf{K}^{A,m}_{\ell} &= \mathbf{H}^{A,m}_{\ell}\mathbf{W}^{A,K}_{\ell,g(m)},\\
        \mathbf{V}^{A,m}_{\ell} &= \mathbf{H}^{A,m}_{\ell}\mathbf{W}^{A,V}_{\ell,g(m)}.
    \end{aligned}
    \label{eq:adapterkv}
\end{equation}
Both projection matrices have shape $d_{\ell}\times d_h$, where $d_h$ is the per-head width, so each projected key or value matrix has shape $L_{\ell,m}\times d_h$. Controls assigned to the same template share its projection parameters; different templates have separate parameters. Equation~\eqref{eq:adapterkv} describes the linear projections before any backbone-specific key normalization or positional transformation. The cache and attention equations use the resulting attention-ready keys and values, retaining the same symbols for brevity. The cache for control $m$ is
\begin{equation}
    \begin{aligned}
        \mathcal{C}^{(m)}_t &= A_{\phi_{g(m)}}(\mathbf{z}_{c,m};\boldsymbol{\eta}_t)\\
        &= \{(\mathbf{K}^{A,m}_{\ell},\mathbf{V}^{A,m}_{\ell})\}_{\ell\in\mathcal{J}},
    \end{aligned}
    \label{eq:cache}
\end{equation}
where $\mathcal{J}$ is the set of injection blocks and $\boldsymbol{\eta}_t$ contains the auxiliary template inputs, including the time embedding and any text context. Because the control inputs are noise-free, the template timestep is fixed at the clean-data endpoint, $1$, and its time embedding is evaluated at this fixed value rather than at the current generation timestep $t$. The remaining template inputs are also held fixed during generation. Consequently, $\boldsymbol{\eta}_t=\boldsymbol{\eta}_1$ and $\mathcal{C}^{(m)}_t=\mathcal{C}^{(m)}_1$ throughout inference: the template outputs do not change across sampling steps. Each active control therefore requires only one template forward pass before iterative generation, after which its layer-wise KV cache is reused at every step. This avoids repeated template computation and substantially reduces the computational overhead of audio conditioning.

\subsection{KV Injection}

The conditioned DiT predicts a latent velocity field:
\begin{equation}
    \widehat{\mathbf{v}}_t
       =F_\theta(\mathbf{z}_t,t,\mathbf{h}_y;\mathcal{C}_t).
    \label{eq:predictor}
\end{equation}
This velocity parameterization is optimized using the flow-matching objective in Section~\ref{sec:training}.

At each selected block, the control keys and values are concatenated with those of the generation branch. The fusion operation uses scaled dot-product attention~\cite{transformer}. Additional reference context is conceptually related to attention-prefix conditioning~\cite{prefix}, although it is computed from the input audio rather than a fixed prefix. For one head, the augmented keys are
\begin{equation}
    \widetilde{\mathbf{K}}_{\ell}
       =[\mathbf{K}^{G}_{\ell};\mathbf{K}^{A}_{\ell}],
    \label{eq:augmented-keys}
\end{equation}
and the augmented values are
\begin{equation}
    \widetilde{\mathbf{V}}_{\ell}
       =[\mathbf{V}^{G}_{\ell};\mathbf{V}^{A}_{\ell}],
    \label{eq:augmented-values}
\end{equation}
where the superscript $G$ denotes the generation branch and $A$ denotes the combined control memory. The resulting attention output is
\begin{equation}
    \mathbf{O}^{G}_{\ell}
       =\operatorname{softmax}\!\left(
          \frac{\mathbf{Q}^{G}_{\ell}\widetilde{\mathbf{K}}_{\ell}^{\top}}
               {\sqrt{d_h}}
         \right)\widetilde{\mathbf{V}}_{\ell}.
    \label{eq:attention}
\end{equation}
Keys and values are concatenated along the token dimension, and softmax is applied over all generation and control keys. Only generation queries produce outputs; multi-head aggregation follows the backbone.

Positional encoding must reflect the semantics of each condition. Beats, vocals, accompaniment, and prosody retain temporal correspondence with the generated sequence, whereas Reference preserves the internal ordering of an excerpt without assigning its events to absolute output positions. Control and generation sequences may therefore differ in length.

\subsection{Audio Conditioning}
\label{sec:controls}
\label{sec:data}

We construct control-target pairs from a private dataset of approximately 60k music recordings. Beats, vocals, accompaniment, and prosody are temporally aligned with the target, whereas Reference provides global context from the same recording. With text annotations $y_i$ and active control subsets $\mathcal{S}_i$, the training dataset is
\begin{equation}
    \mathcal{D}
       =\{(\mathbf{x}_i,y_i,\mathcal{S}_i,\boldsymbol{c}_{i,\mathcal{S}_i})\}_{i=1}^{N},
    \label{eq:dataset}
\end{equation}
where $N$ is the number of constructed examples and $\boldsymbol{c}_{i,\mathcal{S}_i}=\{\mathbf{c}_{i,m}\}_{m\in\mathcal{S}_i}$ contains their control waveforms.

\paragraph{Beats.}
We estimate beat timestamps using Essentia's implementation of the Degara tracker~\cite{essentia,degarabeat} and render them as exponentially decaying sinusoidal pulses:
\begin{equation}
    b(s)=\sum_{k=1}^{N_b}h_{\mathrm{beat}}(s-\tau_k),\qquad
    c_{\mathrm{beats}}(s)=\frac{b(s)}{\|b\|_{\infty}},
    \label{eq:beats}
\end{equation}
where $N_b$ is the beat count, $\tau_k$ is the $k$th timestamp, and $h_{\mathrm{beat}}$ is a 10\,ms pulse with a 1\,kHz carrier, zero outside its support. The waveform spans $0\leq s<T_{\mathbf{x}}$, with $T_{\mathbf{x}}$ the source duration, and $\|b\|_\infty$ is its peak absolute amplitude. Unlike a drum stem, this synthetic track conveys beat timing without source timbre or rhythmic subdivisions.

\paragraph{Vocals and accompaniment.}
Demucs~\cite{demucs} separates each recording into vocal and accompaniment estimates, which serve as the corresponding control inputs. Vocals guide melody and performance and may retain lyrics and singer identity; accompaniment supplies instrumental harmony, rhythm, and arrangement.

\paragraph{Prosody.}
We resynthesize the channel-averaged vocal signal $v[n]$ using a sinusoidal carrier to retain pitch and timing while reducing phonetic and timbral detail. Rectification followed by fourth-order 30\,Hz and second-order 80\,Hz Butterworth low-pass filtering yields the normalized envelope:
\begin{equation}
    \begin{aligned}
        \widetilde{e}[n] &= \mathcal{F}^{(2)}_{80}\!\left(
            \mathcal{F}^{(4)}_{30}(|v|)\right)[n],\\
        e[n] &= \frac{\widetilde{e}[n]}{\max_k\widetilde{e}[k]+\epsilon}.
    \end{aligned}
    \label{eq:prosody-envelope}
\end{equation}
Here, $\mathcal{F}^{(p)}_{f_c}$ applies a filter of one-pass order $p$ and cutoff $f_c$ in both directions, giving zero phase and doubling its effective order; $\epsilon>0$ stabilizes normalization. Filtering may produce negative samples, and no additional clipping is specified.

We estimate fundamental frequency with pYIN~\cite{pyin} over 65--1000\,Hz using a hop of $H=512$ samples. Missing estimates are forward-filled, leading gaps are filled with the first valid estimate, and the trajectory is linearly interpolated to the sample grid with constant boundary extension. At least one valid estimate is required. For an interpolated frequency $f[n]$ in hertz and a processing sample rate $f_s$, the carrier is
\begin{equation}
    \varphi[n]=\frac{2\pi}{f_s}\sum_{k=0}^{n}f[k],
    \qquad
    r[n]=e[n]\sin\!\left(\varphi[n]\right),
    \label{eq:prosody-carrier}
\end{equation}
with initial phase $\varphi[-1]=0$. No voicing mask is applied; the envelope controls amplitude during intervals with imputed pitch. Peak scaling approximately matches the peak amplitude of the vocal reference:
\begin{equation}
    c_{\mathrm{prosody}}[n]
       =r[n]\frac{\max_k|v[k]|}{\max_k|r[k]|+\epsilon}.
    \label{eq:prosody}
\end{equation}
The monaural result is replicated across output channels. This transformation reduces, but does not guarantee removal of, linguistic content or vocal identity.

\paragraph{Reference.}
We select the loudest contiguous $\Delta=10\,\mathrm{s}$ excerpt of the mixed recording to guide style, timbre, and production without prescribing an output timeline. For $T_{\mathbf{x}}\geq\Delta$, let $\mathcal{T}_{\mathbf{x}}$ be the nonempty set of candidate start times on the sample grid within $[0,T_{\mathbf{x}}-\Delta]$:
\begin{align}
    \tau^{\star}
       &\in\operatorname*{arg\,max}_{\tau\in\mathcal{T}_{\mathbf{x}}}
          \operatorname{Loudness}\!\left(\mathbf{x}[\tau:\tau+\Delta]\right),\\
    \mathbf{c}_{\mathrm{reference}}
       &=\mathbf{x}[\tau^{\star}:\tau^{\star}+\Delta].
    \label{eq:reference-selection}
\end{align}
The slice $\mathbf{x}[\tau:\tau+\Delta]$ denotes samples in $[\tau,\tau+\Delta)$. Candidate windows are scored without independent gain normalization; the loudness measure and candidate spacing are preprocessing settings.

\paragraph{Joint conditioning and alignment.}
Each active control is encoded separately. At block $\ell$, its keys and values are concatenated in a fixed order as $\mathbf{K}^{A}_{\ell}=[\mathbf{K}^{A,m}_{\ell}]_{m\in\mathcal{S}}$ and $\mathbf{V}^{A}_{\ell}=[\mathbf{V}^{A,m}_{\ell}]_{m\in\mathcal{S}}$, forming $\mathcal{C}_t=\{(\mathbf{K}^{A}_{\ell},\mathbf{V}^{A}_{\ell})\}_{\ell\in\mathcal{J}}$ for Equations~\eqref{eq:augmented-keys} and~\eqref{eq:augmented-values}. Composition thus combines attention memories, not waveforms. For example, jointly enabling Accompaniment and Prosody allows the model to generate a cover of a song with a user-specified vocal timbre while following its instrumental structure and vocal melody. Combining different controls supports a broader range of music-generation applications.

\subsection{Training Formulation and Inference}
\label{sec:training}

We formulate adapter optimization as conditional flow matching~\cite{flowmatching}: a neural vector field is trained to predict the velocity along a prescribed conditional probability path. Given a clean audio latent $\mathbf{z}_1=E_\omega(\mathbf{x})$,
we independently sample Gaussian noise
$\mathbf{z}_0\sim\mathcal{N}(\mathbf{0},\mathbf{I})$
and a training time $t\sim p(t)$,
where $p$ is a probability distribution on $[0,1]$ induced by the backbone's training timestep schedule expressed in our noise-to-data time convention. We adopt the linear interpolation used in straight-path flow formulations~\cite{rectifiedflow}.
\begin{equation}
    \mathbf{z}_t=(1-t)\mathbf{z}_0+t\mathbf{z}_1.
    \label{eq:flow-path}
\end{equation}
Differentiating with respect to this flow time while holding the sampled endpoints fixed gives the conditional target velocity
\begin{equation}
    \mathbf{u}^{\star}_t
       =\frac{d\mathbf{z}_t}{dt}
       =\mathbf{z}_1-\mathbf{z}_0.
    \label{eq:target-velocity}
\end{equation}
The conditional flow-matching objective~\cite{flowmatching} minimizes squared error between the predicted and target velocities.
\begin{equation}
    \mathcal{L}_{\mathrm{CFM}}(\phi)=\mathbb{E}_{\mathcal{D},\,t,\,\mathbf{z}_0}
    \left[w(t)\left\|\widehat{\mathbf{v}}_t(\phi)-\mathbf{u}^{\star}_t\right\|_2^2\right].
    \label{eq:loss}
\end{equation}
Here, $\widehat{\mathbf{v}}_t(\phi)$ is the prediction in Equation~\eqref{eq:predictor}, with dependence on $\phi$ through the injected cache. The expectation is taken over $(\mathbf{x},y,\mathcal{S},\boldsymbol{c}_{\mathcal{S}})\sim\mathcal{D}$, scheduler-sampled times, and independent Gaussian noise. The nonnegative loss weight $w(t)$ is specified separately from the sampling distribution $p$.

In template-only optimization, the text encoder, VAE, and generation
DiT remain frozen, while gradients propagate through the injected
key-value tensors to update the selected template parameters.
At inference, the separately parameterized Control, Prosody, and
Reference modules can be combined through their attention caches.

The terminal state is decoded as $\widehat{\mathbf{x}}=D_\omega(\widehat{\mathbf{z}}_1)$. At inference, the template timestep is fixed at $1$, and the other template inputs are held constant. The control caches are computed once before integration and reused throughout all sampling steps; only the generation DiT is evaluated iteratively. Auxiliary planning and lyric-conditioning modules are retained when present in the backbone.

\section{Experiments}
\label{sec:experiments}

DiffSynth-Music uses ACE-Step-1.5-XL-SFT~\citep{acestep15} as its backbone, with the audio-conditioned template modules optimized while the generation backbone remains frozen. We evaluate DiffSynth-Music on Mandarin and English song generation and compare it with representative music-generation systems. The evaluation focuses on adherence to audio-derived control signals, including beats, vocals, accompaniment, prosody, and reference audio. We also assess acoustic quality, lyric fidelity, and text--music alignment to examine whether adding controllable generation capabilities preserves general music-generation quality.

\begin{table*}[!t]
\centering
\small
\setlength{\tabcolsep}{3pt}
\caption{Control adherence. Arrows indicate the preferred direction, and bold denotes the best reported value in each column. The DiffSynth-Music row combines results from separate single-control configurations, each using the control indicated by its column group.}
\label{tab:control-results}
\begin{tabular*}{\textwidth}{@{\extracolsep{\fill}}lrrrrrrr@{}}
\toprule
 & \multicolumn{2}{c}{Beats} & \multicolumn{2}{c}{Vocals} & \multicolumn{1}{c}{Accompaniment} & \multicolumn{1}{c}{Prosody} & \multicolumn{1}{c}{Reference} \\
\cmidrule(lr){2-3}\cmidrule(lr){4-5}\cmidrule(lr){6-6}\cmidrule(lr){7-7}\cmidrule(lr){8-8}
Model & Beat-F1$\uparrow$ & Cemgil$\uparrow$ & V-MSE$\downarrow$ & PER$\downarrow$ & A-MSE$\downarrow$ & Pitch$_{50}\uparrow$ & MuLan-A$\uparrow$ \\
\midrule
ACE-Step-1.5-XL-SFT & 0.3031 & 0.1983 & 0.0158 & 0.1928 & 0.0221 & 0.0366 & 0.6836 \\
DiffRhythm-2 & 0.3049 & 0.1998 & 0.0119 & 0.3264 & 0.0204 & 0.0314 & 0.6820 \\
HeartMuLa-3B & 0.3012 & 0.1956 & 0.0182 & 0.2626 & 0.0320 & 0.0347 & 0.5712 \\
MiniMax-Music3 & 0.2921 & 0.1923 & 0.0160 & 0.2663 & 0.0237 & 0.0402 & 0.6566 \\
LeVo-2-Large & 0.2892 & 0.1910 & 0.0165 & 0.2707 & 0.0281 & 0.0349 & 0.5144 \\
\midrule
DiffSynth-Music & \textbf{0.8496} & \textbf{0.8088} & \textbf{0.0006} & \textbf{0.1451} & \textbf{0.0023} & \textbf{0.4671} & \textbf{0.8088} \\
\bottomrule
\end{tabular*}
\end{table*}

\begin{table*}[!t]
\centering
\small
\setlength{\tabcolsep}{4pt}
\caption{Automatic music-quality and instruction-following scores. Higher is better for every metric. CE, CU, PC, and PQ are AudioBox-Aesthetics scores, and MuLan-T evaluates text--music alignment.}
\label{tab:quality-results}
\begin{tabular*}{\textwidth}{@{\extracolsep{\fill}}lrrrrr@{}}
\toprule
 & \multicolumn{4}{c}{AudioBox-Aesthetics} & \multicolumn{1}{c}{Instruction following} \\
\cmidrule(lr){2-5}\cmidrule(lr){6-6}
Model & CE$\uparrow$ & CU$\uparrow$ & PC$\uparrow$ & PQ$\uparrow$ & MuLan-T$\uparrow$ \\
\midrule
ACE-Step-1.5-XL-SFT & 7.3748 & 7.6975 & 6.4335 & 8.1016 & 0.4024 \\
DiffRhythm-2 & 7.2611 & 7.5567 & 6.0345 & 8.0152 & 0.4565 \\
HeartMuLa-3B & 7.6634 & 7.8523 & 6.2728 & 8.2712 & 0.3757 \\
MiniMax-Music3 & 7.3914 & 7.6760 & 6.2354 & 8.0465 & 0.3604 \\
LeVo-2-Large & 7.7488 & 7.9951 & 6.7954 & 8.4514 & 0.2383 \\
\midrule
DiffSynth-Music (Beats) & 7.4484 & 7.8041 & 6.0063 & 8.2155 & 0.3810 \\
DiffSynth-Music (Vocals) & 7.1020 & 7.4263 & 6.1119 & 8.0604 & 0.3452 \\
DiffSynth-Music (Accompaniment) & 7.3535 & 7.5773 & 6.4119 & 8.0555 & 0.3485 \\
DiffSynth-Music (Prosody) & 7.2742 & 7.6471 & 6.1338 & 8.0704 & 0.3625 \\
DiffSynth-Music (Reference) & 7.4436 & 7.7328 & 6.0139 & 8.2418 & 0.3900 \\
\bottomrule
\end{tabular*}
\end{table*}

\subsection{Experimental Setup}

\paragraph{Dataset.}
We collect 100 test music samples: 50 with English lyrics from MUSDB18-HQ~\citep{musdb18hq} and 50 with Mandarin lyrics from our private dataset. Each sample is annotated with a descriptive prompt and lyrics using Qwen3-Omni~\citep{qwen3omni}. For each recording, we extract beats, vocals, accompaniment, prosody, and a reference excerpt as conditioning inputs, following the signal-construction procedures in Section~\ref{sec:data}. These sample-specific conditions provide the targets for evaluating controllability.

\paragraph{Parameters.}
For DiffSynth-Music, we use a classifier-free guidance (CFG)~\citep{cfg} scale of 4 and 50 inference steps. The generation duration matches the duration of the corresponding test sample.

\paragraph{Baselines.}
We compare against ACE-Step-1.5-XL-SFT~\citep{acestep15}, DiffRhythm-2~\citep{diffrhythm}, HeartMuLa-3B~\citep{heartmula}, MiniMax-Music3~\citep{minimax}, and LeVo-2-Large~\citep{levo2}. DiffSynth-Music is built on ACE-Step-1.5-XL-SFT, which does not provide the audio-control capabilities evaluated here. This backbone baseline allows us to examine the effect of adding audio-conditioned templates, while the other systems provide broader comparisons of general song-generation quality.

\subsection{Evaluation Metrics}
\label{sec:metrics}

\paragraph{Beat consistency.}
We use Beat-F1 and Cemgil to evaluate agreement between the input beat sequence and beats extracted from the generated music, following standardized music information retrieval evaluation procedures~\citep{mireval}. Beat-F1 measures beat-position matches within a temporal tolerance, whereas Cemgil uses a Gaussian timing score to quantify deviations from target beats. Higher values indicate better rhythmic alignment.

\paragraph{Vocal control.}
Under vocal conditioning, V-MSE measures the mean squared error between the input vocals and vocals separated from the generated music, using the same signal representation and temporal sampling. Lower values indicate closer vocal reconstruction. We use phoneme error rate (PER)~\citep{heartmula} to evaluate lyric fidelity under vocal conditioning. We separate generated vocals, transcribe them, and compare the resulting phoneme sequence with the input lyrics. PER is the total number of phoneme substitutions, deletions, and insertions divided by the number of reference phonemes. Lower PER indicates better lyric fidelity.

\paragraph{Accompaniment control.}
Under accompaniment conditioning, A-MSE measures the mean squared error between the input accompaniment and accompaniment separated from the generated music, using the same signal representation and temporal sampling. Lower values indicate closer accompaniment reconstruction.

\paragraph{Prosody consistency.}
We use Pitch$_{50}$ to evaluate vocal pitch adherence to the prosody input. Vocal fundamental frequency is estimated with pYIN~\citep{pyin} and compared with the target pitch trajectory on the same timeline. The score is the proportion of target-voiced frames for which the generated vocal is voiced and its pitch lies within 50 cents of the target. Higher values indicate better melodic target following.

\paragraph{Reference-audio consistency.}
MuLan-A measures cosine similarity between the generated audio and the corresponding reference excerpt in the MuQ-MuLan embedding space~\citep{muq}. A higher similarity score indicates closer alignment with the reference audio. Unlike pointwise control-signal MSE, this metric evaluates global audio resemblance without requiring temporal correspondence.

\paragraph{General music-generation quality.}
We use the following metrics to assess whether introducing controllable generation capabilities preserves acoustic quality and text--music alignment, rather than treating control adherence alone as evidence of generation quality.
\begin{itemize}
    \item \textbf{AudioBox-Aesthetics.} AudioBox-Aesthetics~\citep{audiobox} provides four automatic scores: Content-Enjoyment (CE), Content-Usefulness (CU), Production-Complexity (PC), and Production-Quality (PQ). These characterize the appeal and usefulness of the generated content, production complexity, and technical quality, respectively. They serve as proxies for listening quality rather than human listening judgments.
    \item \textbf{Instruction following.} We use MuLan-T, the cosine similarity between the generated audio and its annotated prompt in MuQ-MuLan~\citep{muq}, to evaluate adherence to textual instructions. Higher values indicate stronger text--music alignment.
\end{itemize}

\subsection{Experimental Results}
\label{sec:results}

Tables~\ref{tab:control-results} and~\ref{tab:quality-results} report control adherence and general generation quality, respectively. Each DiffSynth-Music configuration uses one control type. Lower values are better for MSE and PER; higher values are better for all other metrics.

\paragraph{Control adherence.}
DiffSynth-Music achieves the best reported value on all seven control metrics. Beat conditioning raises Beat-F1 from 0.3031 to 0.8496 and Cemgil from 0.1983 to 0.8088 relative to the frozen backbone, indicating substantially closer rhythmic alignment. Vocal conditioning reduces V-MSE from 0.0158 to 0.0006 and PER from 0.1928 to 0.1451, while accompaniment conditioning reduces A-MSE from 0.0221 to 0.0023. The simultaneous improvement in vocal reconstruction and PER suggests that vocal conditioning supports both target adherence and lyric fidelity on the evaluated songs. Prosody conditioning increases Pitch$_{50}$ from 0.0366 to 0.4671, also exceeding the strongest baseline score of 0.0402 from MiniMax-Music3. Reference conditioning improves MuLan-A from 0.6836 to 0.8088. These results support the effectiveness of pitch-and-timing guidance and global reference conditioning, respectively.

\paragraph{Music quality and instruction following.}
Table~\ref{tab:quality-results} shows that DiffSynth-Music retains music quality and instruction-following performance broadly comparable to those of the evaluated base models while adding audio control. Its PQ scores range from 8.0555 to 8.2418, close to the backbone's 8.1016, and its MuLan-T scores of 0.3452--0.3900 fall within the baseline range of 0.2383--0.4565. Beats and Reference also improve CE, CU, and PQ over the backbone. Although some aesthetics scores and all MuLan-T scores are lower than the backbone's, the overall results suggest that controllability is gained while largely retaining these generation capabilities.

\FloatBarrier

\section{Conclusion}
\label{sec:conclusion}

We presented DiffSynth-Music, which extends the frozen ACE-Step-1.5-XL-SFT backbone with three audio-conditioned templates: Control, Prosody, and Reference. Trained with conditional flow matching, these templates support five composable control types through layer-wise KV injection, with control memories computed once and reused throughout sampling. Experiments on Mandarin and English songs demonstrate improved control adherence and vocal-conditioned lyric fidelity, while automatic music-quality and instruction-following scores remain broadly comparable to those of the evaluated base models despite some decreases relative to the backbone. We release the template models to support research and creative applications. Future work will evaluate joint conditioning and musical quality through human listening studies.

\section*{Acknowledgments}

This manuscript was prepared with assistance from GPT for grammar correction and language polishing.

\bibliographystyle{unsrtnat}
\begingroup
\small
\raggedright
\bibliography{references}

@article{musicgen,
  title={Simple and controllable music generation},
  author={Copet, Jade and Kreuk, Felix and Gat, Itai and Remez, Tal and Kant, David and Synnaeve, Gabriel and Adi, Yossi and D{\'e}fossez, Alexandre},
  journal={Advances in neural information processing systems},
  volume={36},
  pages={47704--47720},
  year={2023}
}

@inproceedings{stableaudio,
  title={Stable audio open},
  author={Evans, Zach and Parker, Julian D and Carr, CJ and Zukowski, Zack and Taylor, Josiah and Pons, Jordi},
  booktitle={ICASSP 2025-2025 IEEE International Conference on Acoustics, Speech and Signal Processing (ICASSP)},
  pages={1--5},
  year={2025},
  organization={IEEE}
}

@article{acestep,
  title={Ace-step: A step towards music generation foundation model},
  author={Gong, Junmin and Zhao, Sean and Wang, Sen and Xu, Shengyuan and Guo, Joe},
  journal={arXiv preprint arXiv:2506.00045},
  year={2025}
}

@article{acestep15,
  title={Ace-step 1.5: Pushing the boundaries of open-source music generation},
  author={Gong, Junmin and Song, Yulin and Zhao, Wenxiao and Wang, Sen and Xu, Shengyuan and Guo, Jing and Yang, Xuerui},
  journal={arXiv preprint arXiv:2602.00744},
  year={2026}
}

@misc{minimax,
  author = {{MiniMax}},
  title = {{MiniMax-Music3}: Official Repository},
  year = {2026},
  url = {https://github.com/MiniMax-AI/MiniMax-Music3},
  note = {Software and model documentation; accessed September 6, 2026}
}

@inproceedings{yue,
  title={{YuE}: Scaling open foundation models for long-form music generation},
  author={Yuan, Ruibin and Lin, Hanfeng and Guo, Shuyue and Zhang, Ge and Pan, Jiahao and Zang, Yongyi and Liu, Haohe and Liang, Yiming and Ma, Wenye and Du, Xingjian and others},
  booktitle={International Conference on Learning Representations},
  year={2026},
  url={https://openreview.net/forum?id=hZy6YG2Ij8}
}

@inproceedings{dit,
  title={Scalable diffusion models with transformers},
  author={Peebles, William and Xie, Saining},
  booktitle={2023 IEEE/CVF International Conference on Computer Vision (ICCV)},
  pages={4172--4182},
  year={2023},
  organization={IEEE}
}

@inproceedings{controlnet,
  title={Adding conditional control to text-to-image diffusion models},
  author={Zhang, Lvmin and Rao, Anyi and Agrawala, Maneesh},
  booktitle={2023 IEEE/CVF International Conference on Computer Vision (ICCV)},
  pages={3813--3824},
  year={2023},
  organization={IEEE}
}

@inproceedings{prefix,
  title={Prefix-tuning: Optimizing continuous prompts for generation},
  author={Li, Xiang Lisa and Liang, Percy},
  booktitle={Proceedings of the 59th annual meeting of the association for computational linguistics and the 11th international joint conference on natural language processing (volume 1: Long papers)},
  pages={4582--4597},
  year={2021}
}

@inproceedings{essentia,
  title={Essentia: An audio analysis library for music information retrieval},
  author={Bogdanov, Dmitry and Wack, Nicolas and G{\'o}mez Guti{\'e}rrez, Emilia and Gulati, Sankalp and Herrera Boyer, Perfecto and Mayor, Oscar and Roma Trepat, Gerard and Salamon, Justin and Zapata Gonz{\'a}lez, Jos{\'e} Ricardo and Serra, Xavier},
  booktitle={Britto A, Gouyon F, Dixon S, editors. 14th Conference of the International Society for Music Information Retrieval (ISMIR); 2013 Nov 4-8; Curitiba, Brazil.[place unknown]: ISMIR; 2013. p. 493-8.},
  year={2013},
  organization={International Society for Music Information Retrieval (ISMIR)}
}

@inproceedings{demucs,
  title={Hybrid transformers for music source separation},
  author={Rouard, Simon and Massa, Francisco and D{\'e}fossez, Alexandre},
  booktitle={ICASSP 2023-2023 IEEE International Conference on Acoustics, Speech and Signal Processing (ICASSP)},
  pages={1--5},
  year={2023},
  organization={IEEE}
}

@article{musiccontrolnet,
  title={Music controlnet: Multiple time-varying controls for music generation},
  author={Wu, Shih-Lun and Donahue, Chris and Watanabe, Shinji and Bryan, Nicholas J},
  journal={IEEE/ACM Transactions on Audio, Speech, and Language Processing},
  volume={32},
  pages={2692--2703},
  year={2024},
  publisher={IEEE}
}

@article{jasco,
  title={Joint audio and symbolic conditioning for temporally controlled text-to-music generation},
  author={Tal, Or and Ziv, Alon and Gat, Itai and Kreuk, Felix and Adi, Yossi},
  journal={arXiv preprint arXiv:2406.10970},
  year={2024}
}

@article{vae,
  title={Auto-encoding variational bayes},
  author={Kingma, Diederik P and Welling, Max},
  journal={arXiv preprint arXiv:1312.6114},
  year={2013}
}

@article{ddpm,
  title={Denoising diffusion probabilistic models},
  author={Ho, Jonathan and Jain, Ajay and Abbeel, Pieter},
  journal={Advances in neural information processing systems},
  volume={33},
  pages={6840--6851},
  year={2020}
}

@inproceedings{latentdiffusion,
  title={High-resolution image synthesis with latent diffusion models},
  author={Rombach, Robin and Blattmann, Andreas and Lorenz, Dominik and Esser, Patrick and Ommer, Bj{\"o}rn},
  booktitle={2022 IEEE/CVF conference on computer vision and pattern recognition (CVPR)},
  pages={10674--10685},
  year={2022},
  organization={ieee}
}

@article{flowmatching,
  title={Flow matching for generative modeling},
  author={Lipman, Yaron and Chen, Ricky TQ and Ben-Hamu, Heli and Nickel, Maximilian and Le, Matt},
  journal={arXiv preprint arXiv:2210.02747},
  year={2022}
}

@article{transformer,
  title={Attention is all you need},
  author={Vaswani, Ashish and Shazeer, Noam and Parmar, Niki and Uszkoreit, Jakob and Jones, Llion and Gomez, Aidan N and Kaiser, {\L}ukasz and Polosukhin, Illia},
  journal={Advances in neural information processing systems},
  volume={30},
  year={2017}
}

@inproceedings{t2iadapter,
  title={T2i-adapter: Learning adapters to dig out more controllable ability for text-to-image diffusion models},
  author={Mou, Chong and Wang, Xintao and Xie, Liangbin and Wu, Yanze and Zhang, Jian and Qi, Zhongang and Shan, Ying},
  booktitle={Proceedings of the AAAI conference on artificial intelligence},
  volume={38},
  pages={4296--4304},
  year={2024}
}

@article{lora,
  title={Lora: Low-rank adaptation of large language models},
  author={Hu, Edward J and Shen, Yelong and Wallis, Phillip and Allen-Zhu, Zeyuan and Li, Yuanzhi and Wang, Shean and Wang, Lu and Chen, Weizhu},
  journal={arXiv preprint arXiv:2106.09685},
  year={2021}
}

@article{multifeaturebeat,
  title={Multi-feature beat tracking},
  author={Zapata, Jos{\'e} R and Davies, Matthew EP and G{\'o}mez, Emilia},
  journal={IEEE/ACM Transactions on Audio, Speech, and Language Processing},
  volume={22},
  number={4},
  pages={816--825},
  year={2014},
  publisher={IEEE}
}

@article{degarabeat,
  title={Reliability-informed beat tracking of musical signals},
  author={Degara, Norberto and R{\'u}a, Enrique Argones and Pena, Antonio and Torres-Guijarro, Soledad and Davies, Matthew EP and Plumbley, Mark D},
  journal={IEEE Transactions on Audio, Speech, and Language Processing},
  volume={20},
  number={1},
  pages={290--301},
  year={2012},
  publisher={IEEE}
}

@article{rectifiedflow,
  title={Flow straight and fast: Learning to generate and transfer data with rectified flow},
  author={Liu, Xingchao and Gong, Chengyue and Liu, Qiang},
  journal={arXiv preprint arXiv:2209.03003},
  year={2022}
}

@inproceedings{pyin,
  title={pYIN: A fundamental frequency estimator using probabilistic threshold distributions},
  author={Mauch, Matthias and Dixon, Simon},
  booktitle={2014 ieee international conference on acoustics, speech and signal processing (icassp)},
  pages={659--663},
  year={2014},
  organization={IEEE}
}

@article{heartmula,
  title={Heartmula: A family of open sourced music foundation models},
  author={Yang, Dongchao and Xie, Yuxin and Yin, Yuguo and Wang, Zheyu and Yi, Xiaoyu and Zhu, Gongxi and Weng, Xiaolong and Xiong, Zihan and Ma, Yingzhe and Cong, Dading and others},
  journal={arXiv preprint arXiv:2601.10547},
  year={2026}
}

@article{diffrhythm,
  title={Diffrhythm 2: Efficient and high fidelity song generation via block flow matching},
  author={Jiang, Yuepeng and Chen, Huakang and Ning, Ziqian and Yao, Jixun and Han, Zerui and Wu, Di and Meng, Meng and Luan, Jian and Fu, Zhonghua and Xie, Lei},
  journal={arXiv preprint arXiv:2510.22950},
  year={2025}
}

@inproceedings{audiobox,
  title={Meta {Audiobox Aesthetics}: Unified automatic quality assessment for speech, music, and sound},
  author={Tjandra, Andros and Wu, Yi-Chiao and Guo, Baishan and Hoffman, John and Ellis, Brian and Vyas, Apoorv and Shi, Bowen and Chen, Sanyuan and Le, Matt and Zacharov, Nick and others},
  booktitle={2025 IEEE Automatic Speech Recognition and Understanding Workshop (ASRU)},
  pages={1--8},
  year={2025},
  organization={IEEE}
}

@article{muq,
  title={Muq: Self-supervised music representation learning with mel residual vector quantization},
  author={Zhu, Haina and Zhou, Yizhi and Chen, Hangting and Yu, Jianwei and Ma, Ziyang and Gu, Rongzhi and Luo, Yi and Tan, Wei and Chen, Xie},
  journal={IEEE Transactions on Audio, Speech and Language Processing},
  year={2025},
  publisher={IEEE}
}

@article{songbloom,
  title={{SongBloom}: Coherent song generation via interleaved autoregressive sketching and diffusion refinement},
  author={Yang, Chenyu and Wang, Shuai and Chen, Hangting and Tan, Wei and Yu, Jianwei and Li, Haizhou},
  journal={Advances in Neural Information Processing Systems},
  volume={38},
  pages={34256--34277},
  year={2025}
}

@article{levo2,
  title={LeVo 2: Stable and Melodious Song Generation via Hierarchical Representation Modeling and Progressive Post-Training},
  author={Lei, Shun and Zhang, Huaicheng and Wu, Dapeng and Xu, Yaoxun and Zuo, Lishi and Tan, Wei and Chen, Hangting and Li, Guangzheng and Yu, Jianwei and Wu, Zhiyong and others},
  journal={arXiv preprint arXiv:2606.30642},
  year={2026}
}

@article{ipadapter,
  title={Ip-adapter: Text compatible image prompt adapter for text-to-image diffusion models},
  author={Ye, Hu and Zhang, Jun and Liu, Sibo and Han, Xiao and Yang, Wei},
  journal={arXiv preprint arXiv:2308.06721},
  year={2023}
}

@misc{musdb18hq,
  title={{MUSDB18-HQ} -- an uncompressed version of {MUSDB18}},
  author={Rafii, Zafar and Liutkus, Antoine and St{\"o}ter, Fabian-Robert and Mimilakis, Stylianos Ioannis and Bittner, Rachel},
  howpublished={Zenodo},
  year={2019},
  doi={10.5281/zenodo.3338373}
}

@article{qwen3omni,
  title={Qwen3-omni technical report},
  author={Xu, Jin and Guo, Zhifang and Hu, Hangrui and Chu, Yunfei and Wang, Xiong and He, Jinzheng and Wang, Yuxuan and Shi, Xian and He, Ting and Zhu, Xinfa and others},
  journal={arXiv preprint arXiv:2509.17765},
  year={2025}
}

@article{cfg,
  title={Classifier-free diffusion guidance},
  author={Ho, Jonathan and Salimans, Tim},
  journal={arXiv preprint arXiv:2207.12598},
  year={2022}
}

@article{diffusiontemplates,
  title={Diffusion Templates: A Unified Plugin Framework for Controllable Diffusion},
  author={Duan, Zhongjie and Zhang, Hong and Chen, Yingda},
  journal={arXiv preprint arXiv:2604.24351},
  year={2026}
}

@inproceedings{mireval,
  title={{mir\_eval}: A transparent implementation of common {MIR} metrics},
  author={Raffel, Colin and McFee, Brian and Humphrey, Eric J and Salamon, Justin and Nieto, Oriol and Liang, Dawen and Ellis, Daniel P. W.},
  booktitle={Proceedings of the 15th International Society for Music Information Retrieval Conference},
  pages={367--372},
  year={2014}
}
\endgroup
\end{document}